\documentclass[%
 reprint,
superscriptaddress,
 amsmath,amssymb,
 aps,
 longbibliography,
floatfix,
]{revtex4-2}

\usepackage{graphicx}
\usepackage{dcolumn}
\usepackage{bm}
\usepackage{float}
\usepackage[colorlinks=true, citecolor=blue, linkcolor=blue, urlcolor=blue]{hyperref}

\begin{document}

\title{Observation of Erratic Non-Hermitian Skin Effect in Phononic Crystals}

\author{Yujian Yuan}
\thanks{These authors contribute equally to this work.}
\affiliation{Institute of Acoustics, School of Physics Science and Engineering, Tongji University, Shanghai, China.}

\author{Jie Liu}
\thanks{These authors contribute equally to this work.}
\affiliation{Institute of Acoustics, School of Physics Science and Engineering, Tongji University, Shanghai, China.}

\author{He Gao}
\email{hegao@nju.edu.cn}
\affiliation{School of Advanced Manufacturing Engineering, Nanjing University, Suzhou, China.}

\author{Jiamin Guo}
\affiliation{Institute of Acoustics, School of Physics Science and Engineering, Tongji University, Shanghai, China.}

\author{Zhongming Gu}
\email{zhmgu@tongji.edu.cn}
\affiliation{Institute of Acoustics, School of Physics Science and Engineering, Tongji University, Shanghai, China.}

\author{Jie Zhu}
\email{jiezhu@tongji.edu.cn}
\affiliation{Institute of Acoustics, School of Physics Science and Engineering, Tongji University, Shanghai, China.}

\date{\today}

\begin{abstract}
The erratic non-Hermitian skin effect (ENHSE), emerging from the interplay between 
disorders and locally nonreciprocal yet globally reciprocal couplings, has reshaped 
the conventional bulk-boundary correspondence through its disorder-dependent localization 
properties. Here, we experimentally observe the dynamical phenomena of ENHSE in 
phononic crystals with disordered imaginary gauge fields. The erratic localization 
occurs in the bulk independent of the excitation position, with the main and satellite 
peaks precisely located at the local maxima of the cumulative gauge field in accordance 
with random-walk extreme-value statistics. Remarkably, the selective manipulation of 
satellite peaks can be realized by tuning the staggered disorder strengths in a 
dimerized chain. These findings can deepen the understanding of non-Hermitian physics 
and establish a new route for disorder-engineered non-Hermitian wave control.  
\end{abstract}

\maketitle

\textit{Introduction}---In recent years, non-Hermitian physics has garnered increasing attention
within the research community since it transcends the conventional framework of Hermitian
assumption\cite{Ashida20}.  Non-Hermitian systems generally
exhibit complex energy spectra, exceptional points\cite{Ding22,Bergholtz21}, and non-equilibrium
dynamics\cite{Li24,Pyrialakos22}, thereby offering a versatile platform for investigating unexpected
spectral structures as well as wave propagation phenomena in quantum and classical systems. Among them, the non-Hermitian skin effect (NHSE) is a
representative phenomenon\cite{LinR23}, characterized by the accumulation of bulk states at
the boundaries of the system\cite{Li24}. The NHSE invalidates the
conventional Bloch-band description, exhibiting extreme sensitivity to boundary
conditions and prompting the development of generalized Brillouin-zone and
non-Bloch band theories to restore the bulk–boundary correspondence\cite{Yao18,Yokomizo19,Yao18_1}. This
effect has been experimentally observed in various platforms, including
photonic\cite{Zhou23,Wang25c,Sun24}, acoustic\cite{Zhang21,Zhong25,Gu22}, and electrical circuit systems\cite{Zhang23b,Zou21,Liu23}.

NHSE triggered rich interactions with other wave effects and gives rise to a series of important variants, 
such as hybrid topological NHSE\cite{JiangT24,ZouD21}, Floquet-engineered NHSE\cite{ZhangQ26,KeS23}, higher-order NHSE\cite{ZhangX21,KawabataK20}, 
as well as nonlinear NHSE\cite{WangS25,YUCE21} and nonequilibrium NHSE\cite{Gu22,LiZ24}. Disorder also significantly 
affects the spatial and spectral properties of NHSE. In Hermitian systems, disorder leads to 
Anderson localization\cite{Anderson58} through multiple scattering and interference, with wave functions decaying 
exponentially and long-range transport being suppressed. By introducing disorders to the onsite or coupling terms 
of NHSE models, the interplay between nonreciprocity and randomness 
gives rise to unconventional localization behaviors beyond the conventional Anderson paradigm. 
Wang \textit{et al.}\ \cite{Wang25} demonstrated disorder-induced boundary localization in non-Hermitian 
acoustic crystals, where non-Hermitian disorder drives the emergence of NHSE even in systems that are trivial 
in the clean limit, enabling controllable and even bipolar skin localization. Meanwhile, Anderson delocalization 
in strongly coupled disordered non-Hermitian chains has been reported by Jin \textit{et al.}~\cite{JinWW25}, showing that 
interchain coupling can counterintuitively restore delocalization and reintroduce the NHSE even in the strong-disorder regime.

By precisely controlling the non-reciprocal disorder while preserving the global 
reciprocity, one can further enforce bulk eigenstates forming localization peaks 
at erratic positions, a phenomenon referred to as  
ENHSE\cite{Longhi25}. Very recently, experimental observations of erratic non-Hermitian skin 
localization and Floquet ENHSE in undimerized lattices have been reported in acoustic\cite{Zhong26} 
and photonic\cite{Sun26} platforms, respectively. This new form of erratic non-Hermitian skin localization 
manifests as peaks within the bulk, whose positions vary randomly across different 
disorder realizations, revealing a previously unobserved mechanism for wave localization 
in non-Hermitian systems. Unlike conventional standard NHSE or Anderson localization, 
ENHSE exhibits fundamentally distinct features: in contrast to Anderson localization, 
wave functions concentrate at identical random positions within the bulk, forming pronounced 
peaks with spatial profiles that are no longer strictly exponential; in contrast to NHSE, 
ENHSE does not rely on system boundaries, and the peak locations and amplitudes are determined 
by the spatial randomness of disorder. However, the experimental implementation of ENHSE, especially 
the active control of its satellite peaks, has remained elusive due to the difficulty of widely tuning 
nonreciprocal couplings while preserving global reciprocity, which also hinders the exploration of the 
non-Hermitian dynamics of ENHSE in more complex models. 

In this work, we utilize the acoustic platform to systematically investigate the interplay 
between non-Hermitian skin effects and controllable disorder, unveiling the resulting erratic 
localization behavior. By introducing tunable non-reciprocal disorder in a one-dimensional 
non-Hermitian acoustic crystal, we experimentally observe the intriguing localization phenomena 
of ENHSE, reflecting the interplay between spectral structure, boundary conditions, and disorder. 
Moreover, in the dimerized model, the wave function can be selectively localized at the even or 
odd sites, according to the disorder strength of staggered couplings. These experimental results 
not only reveal a novel localization mechanism emerging from the synergy of non-Hermitian couplings 
and disorder, but also provide a feasible route for controlling wave dynamics through engineered 
disorder.

\begin{figure*}[t]
  \centering
  \includegraphics[page=1,scale=0.85]{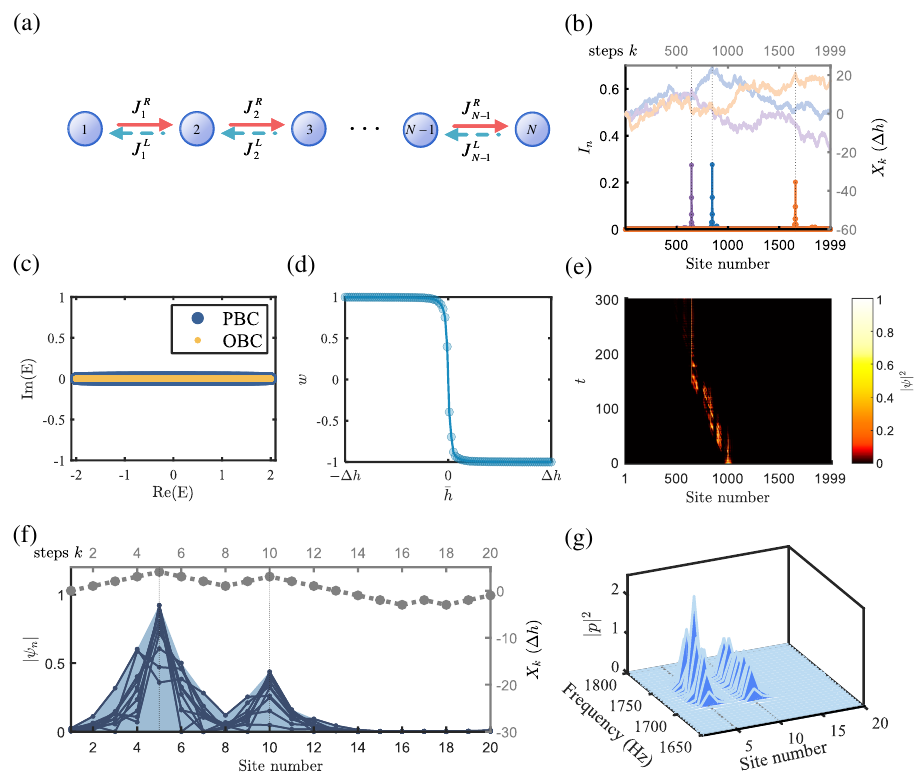}%
  \caption{\label{fig:1}Erratic Non-Hermitian Skin Effect. (a) Schematic diagram of the one-dimensional non-reciprocal Hatano--Nelson model with $N$ sites. (b) Comparison of random-walk distributions and eigenstate distributions under open boundary conditions (OBC) for $N=1999$.  (c)  Eigenenergy spectra under periodic boundary conditions (PBC) and OBC. (d) Winding numbers averaged over 50 randomly generated sequences for different mean values, with lattice size $N=1999$.  (e) Time evolution of a random walk process. (f) Random-walk distributions and eigenstate distributions under OBC for $N=20$. The shaded light-blue area represents the wave packet of the corresponding eigenstates. (g) Numerically simulated response of the model in the acoustic system.}
\end{figure*}

\textit{Model}---We consider a one-dimensional disordered Hatano--Nelson model\cite{Hatano96,Hatano98}, in which the
disorder is introduced through fluctuations of the imaginary gauge field
following the approach proposed in Ref.~\cite{Longhi25}. The system consists of a
lattice of $N$ sites with non-reciprocal hopping, as illustrated in
Fig.~\ref{fig:1}(a). The right- and left-directional hopping amplitudes between
adjacent sites are denoted by $J_n^{R}$ and $J_n^{L}$, respectively. In
the presence of a fluctuating imaginary gauge field, the hopping
amplitudes are parametrized as:
\begin{equation}
J_{n}^{L(R)}=J e^{\mp h_n},
\label{eq:hopping}
\end{equation}
where $h_n$ forms a stochastic sequence $\{h_n\}$ consisting of
independent random variables that share the same probability density
function $f(h)$ with mean $\bar{h}$ and finite variance ${{(\Delta h)}^{2}}$. Such
disorder induces random non-reciprocal hopping and gives rise to
unconventional localization phenomena unique to non-Hermitian systems.

The system is described by the tight-binding Hamiltonian:
\begin{equation}
\hat{H}=\sum\limits_{n=1}^{N-1}{(}J_{n}^{R}\hat{c}_{n+1}^{\dagger }{{\hat{c}}_{n}}+J_{n}^{L}\hat{c}_{n}^{\dagger }{{\hat{c}}_{n+1}})+{{\hat{H}}_{B}},
\label{eq:hamiltonian}
\end{equation}
where $\hat{H}_{B}$ denotes the boundary term. The spatial
localization of eigenstates corresponds to the extreme-value statistics
of the cumulative imaginary gauge field:
\begin{equation}
{{X}_{n}}=\sum\limits_{l=0}^{n-1}{{{h}_{l}}}.
\label{eq:phi}
\end{equation}
In Fig.~\ref{fig:1}(b), we consider a large lattice
with $N=1999$ and compute the spatial eigenstate distribution:
\begin{equation}
{I}_{n}=\frac{1}{N}\sum_{\alpha=1}^{N}\big|\psi_\alpha(n)\big|^2,
\label{eq:profile}
\end{equation}
for three independent realizations of the hopping-disorder sequence ${h_n}$, 
each element drawn from a Bernoulli distribution taking values $\pm \Delta h$ 
with equal probability (0.5 each). We take $\Delta h=0.4$ and $J=1$. 
The quantity ${{I}_{n}}$ characterizes the
averaged spatial weight of all eigenstates. For conventional Anderson
localization, ${{I}_{n}}$ would remain nearly uniform across the sample,
whereas for the standard NHSE it would be strongly localized at the
system boundaries. In contrast, in the erratic regime, ${{I}_{n}}$ develops
pronounced peaks at positions corresponding to the extrema of the
associated random-walk trajectory ${{X}_{n}}$. For each disorder
realization, the peak positions of ${{I}_{n}}$ exhibit a direct correspondence
with the extremal positions of ${{X}_{n}}$, demonstrating that the
localization pattern reflects the extreme-value statistics of the
cumulative imaginary gauge field.

The non-Hermitian hopping asymmetry leads to the single-particle
eigenvalue equation:
\begin{equation}
E{{\psi }_{n}}=J_{n}^{L}{{\psi }_{n+1}}+J_{n-1}^{R}{{\psi }_{n-1}}.
\label{eq:eig}
\end{equation}

Introducing the cumulative gauge field ${{X}_{n}}$ and applying the
non-unitary gauge transformation ${{\psi }_{n}}={{\phi }_{n}} e^{{{X}_{n}}}$\cite{Midya24}, the
cumulative exponential factor absorbs the local asymmetry. Under the
condition that the average non-reciprocity vanishes,
$\bar{h}=0$, the transformed equation becomes:
\begin{equation}
E{{\phi }_{n}}=J({{\phi }_{n+1}}+{{\phi }_{n-1}}),
\label{eq:hermiteig}
\end{equation}
which is formally identical to the eigenvalue equation of a typical
one-dimensional Hermitian chain. Hence, the transformed spectrum
exhibits a Hermitian-like structure for $\bar{h}=0$. This is
further confirmed by the eigenenergy spectra under periodic and open
boundary conditions, as shown in Fig.~\ref{fig:1}(c). The two spectra nearly
coincide, indicating that the spectrum remains insensitive to boundary
conditions in this regime (see Sec.~I of the Supplemental Material~\cite{SM} for additional comparisons between PBC and OBC).

To characterize the topological properties of the non-Hermitian lattice,
we compute the real-space winding number from the single-particle
eigenstates under OBC\cite{Claes21,Wang25,Song18}. The 
real-space winding number is defined as:
\begin{equation}
w(E)=\frac{1}{N}\text{Tr}({\hat{Q}}^{\dagger }[\hat{Q},\hat{X}]),
\label{eq:winding}
\end{equation}
with $\hat{X}$ denoting the position operator along the chain. We
construct a doubled Hermitian Hamiltonian
$\hat{\mathcal{H}}_d=\begin{pmatrix}0 & \hat{H}-E_{\text{ref}} \\ (\hat{H}-E_{\text{ref}})^\dagger & 0\end{pmatrix}$,
where $E_{\text{ref}}$ is a reference energy in the complex plane. This
construction facilitates the computation of the real-space winding
number while preserving the chiral symmetry of the doubled Hamiltonian.
From the doubled Hamiltonian $\hat{\mathcal{H}}_d$, we define a projection
operator $\hat{P}$. This operator projects the full Hilbert space onto the
states with negative eigenvalues of $\hat{\mathcal{H}}_d$, effectively
selecting the occupied subspace. Using $\hat{P}$, we construct the unitary
operator $\hat{Q}_d=1-2\hat{P}$, which encodes the topological properties of the
original Hamiltonian relative to the reference energy $E_{\text{ref}}$.
In block form, $\hat{Q}_d$ can be written as $\begin{pmatrix}0 & \hat{Q} \\ {\hat{Q}}^{\dagger } &
0\end{pmatrix}$, where $\hat{Q}$ acts on the subspace of the original
Hamiltonian.

We evaluate the real-space winding number on a lattice of 
length $N=1999$ and perform an ensemble average over 50 independently 
generated disorder realizations for each value of the mean bias $\bar{h}$, 
thereby approximating the thermodynamic, disorder-averaged response while 
suppressing sample-to-sample fluctuations. As shown in Fig.~\ref{fig:1}(d), 
the averaged winding number depends sensitively on $\bar{h}$ and vanishes at $\bar{h}=0$, 
indicating a trivial point-gap topology.indicating a trivial point-gap topology. Notably,
even when the system exhibits a nontrivial point-gap
topology under OBC for finite $\bar{h}$, the eigenstates remain
primarily localized due to stochastic extreme-value fluctuations
of the cumulative imaginary gauge field, rather than being driven by the system's topology. This demonstrates
that, although the system possesses nontrivial topology,
it does not constitute the dominant mechanism responsible
for localization in the weak-bias regime. As the mean
bias increases, the winding number gradually approaches $\pm1$, signaling the
strengthening of the nontrivial point-gap topology\cite{Yao18,Lin23}. 
In this regime, boundary pumping
progressively becomes more significant, and the system
crosses over toward the conventional NHSE characterized
by edge-localized eigenstates.

To further illustrate the dynamical manifestation of erratic
non-Hermitian skin localization, we study the time evolution of a
single-particle Gaussian wave packet under OBC\cite{LiWan22,Hu23,LiH22}.
In the time-evolution calculation, we adopt the same hopping sequence as that corresponding to the purple curve in Fig.~\ref{fig:1}(b). The
Hamiltonian is constructed using the same randomly generated sequence of
the local imaginary gauge field $\{{{h}_{n}}\}$ as in the previous static calculations.
The initial wave packet is centered at the middle of the lattice with a
width $\sigma =20$ and initial momentum $k_0=0$, given by:
\begin{equation}
\psi_n(t=0) = \frac{1}{(2 \pi \sigma^2)^{1/4}}
\exp\Big[-\frac{(n-n_0)^2}{4\sigma^2} + i k_0 n \Big].
\end{equation}
The time evolution is
computed using the Crank--Nicolson method adapted for non-Hermitian
systems, with a total evolution time ${{t}_{\text{max}}}=300$ and ${{N}_{t}}=500$ discrete time
steps.

The resulting probability distributions $|{{\psi }_{n}}(t){{|}^{2}}$ normalized to their instantaneous
maxima are presented in Fig.~\ref{fig:1}(e). Owing to the use of the same disorder
realization, the spatial positions where the wave packet dynamically
accumulates correspond directly to the extremal positions indicated by
the purple random-walk trajectory in Fig.~\ref{fig:1}(b). As time evolves, the
wave packet progressively localizes near the same extrema predicted by
the cumulative imaginary gauge field. This result reflects the influence of the underlying stochastic hopping 
structure on the dynamical localization of the wave packet.

Importantly, the erratic effect mechanism remains robust even in small systems. Under open boundary conditions, 
a 20-site lattice exhibits similar spatial accumulation of eigenstates. As shown in 
Fig.~\ref{fig:1}(f), the wave-function intensity of the eigenstate distributions (light blue shaded regions) 
aligns closely with the extremal points of the cumulative random-walk trajectory (grey dashed line). The 
largest peak occurs at site 5, with another significant effect at site 10, directly corresponding to 
the maxima of the same hopping-disorder sequence. This demonstrates that the stochastic extreme-value mechanism 
governing erratic skin effect remains valid in experimentally accessible finite lattices.
The numerically simulated response of the model in an acoustic system [see Fig.~\ref{fig:1}(g)] 
reproduces the characteristic irregular effect pattern predicted by the theory. These results provide support 
for the experimental realization of the stochastic extreme-value mechanism governing the erratic skin effect 
in a relatively small acoustic lattice, with all experimental parameters detailed in the Supplementary Material \cite{SM}.

For the ENHSE, the localization behavior is fundamentally distinct from that 
of the conventional non-Hermitian skin effect. In particular, under the condition of vanishing average 
non-reciprocity, $\bar{h}=0$, the Lyapunov exponent becomes zero\cite{Longhi25}, indicating the absence of conventional exponential 
localization. Nevertheless, the eigenstates remain localized due to the stochastic extreme-value mechanism induced 
by the cumulative imaginary gauge field. A further comparison of the Lyapunov spectra of the disordered Hatano--Nelson model for two system sizes, $N=20$ and $N=1999$, 
shows that the two curves are very similar and remain close to zero over a broad frequency window. The weak size dependence indicates that finite-size effects on the 
Lyapunov exponent are already small even for relatively small systems (see Supplemental Material~\cite{SM}, Sec.~II).

\begin{figure}[t]
  \centering
  \hspace*{-0.05\columnwidth}%
  \includegraphics[page=2,scale=0.85]{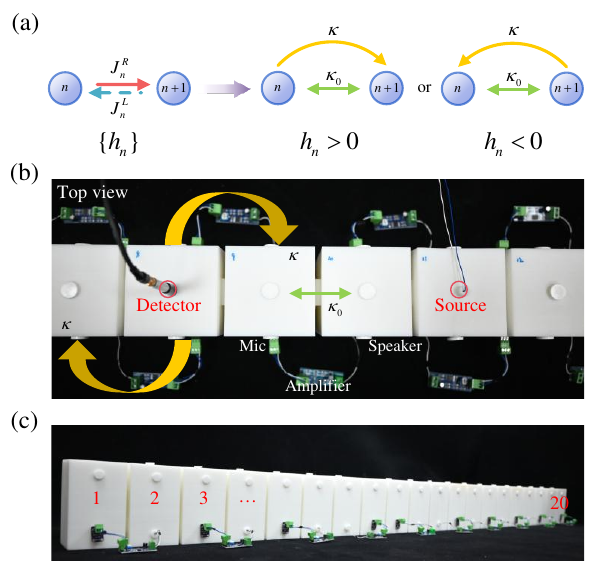}%
  \caption{\label{fig:2}Experimental Implementation. (a) Equivalent Hatano--Nelson model with the same mathematical structure. (b) Top-view photograph of a one-dimensional disordered acoustic crystal. Non-reciprocal coupling is implemented via an amplifier (with DC power supply), a microphone (input), and a speaker (output). Source and detector are used to measure transmission. The green and yellow arrows denote the reciprocal coupling and the unidirectional nonreciprocal coupling, respectively. (c) Photograph of a one-dimensional disordered acoustic crystal with 20 resonators.}
\end{figure}

\textit{Acoustic implementation}---To simplify the experimental realization, we employ an ingenious mathematical 
transformation that the non-reciprocal hoppings $J_{n}^{R}$ and $J_{n}^{L}$ are
mapped onto a fixed reciprocal coupling ${{\kappa}_{0}}$ and a unidirectional
non-reciprocal coupling $\kappa$, defined as:
\begin{equation}
{{\kappa}_{0}}=J{{e}^{-\Delta h}},\quad \kappa =2J\sinh (\Delta h),
 \label{eq:directional}
\end{equation}
with the direction of $\kappa$ between neighboring sites determined by the
sign of ${{h}_{n}}$. The coupling is directed along the positive direction
when ${{h}_{n}}>0$ and along the negative direction when ${{h}_{n}}<0$,
preserving the mathematical structure of the original hoppings illustrated in
Fig.~\ref{fig:2}(a). By randomly assigning the sign of ${h}_{n}$ according to a 
Bernoulli distribution, we determine the direction of the non-reciprocal 
coupling between neighboring sites, thereby directly introducing disorder
into the lattice.

The one-dimensional acoustic lattice consists of resonant cavities that
serve as discrete lattice sites. Adjacent sites are coupled through
connecting tubes that provide the reciprocal coupling ${{\kappa}_{0}}$.
Figure~\ref{fig:2}(b) provides a close-up view of the experimental implementation, highlighting 
the site-resolved cavity arrangement and inter-site coupling pathways. Each cavity 
contains a microphone--speaker pair linked through an external amplifier circuit to 
realize the unidirectional coupling $\kappa$. Cylindrical holes are drilled on the sample 
surface to mount the microphone--speaker pairs for non-reciprocal coupling, while 
additional top openings accommodate the external source and detector used to probe the 
intracavity sound pressure. In the schematic, the green arrow denotes the fixed reciprocal 
coupling $\kappa_{0}$, whereas the yellow arrows mark the single-directional 
coupling $\kappa$. The complete 20-resonator experimental device is presented in 
Fig.~\ref{fig:2}(c).

\begin{figure*}[t]
  \centering
  \includegraphics[page=3,scale=0.85]{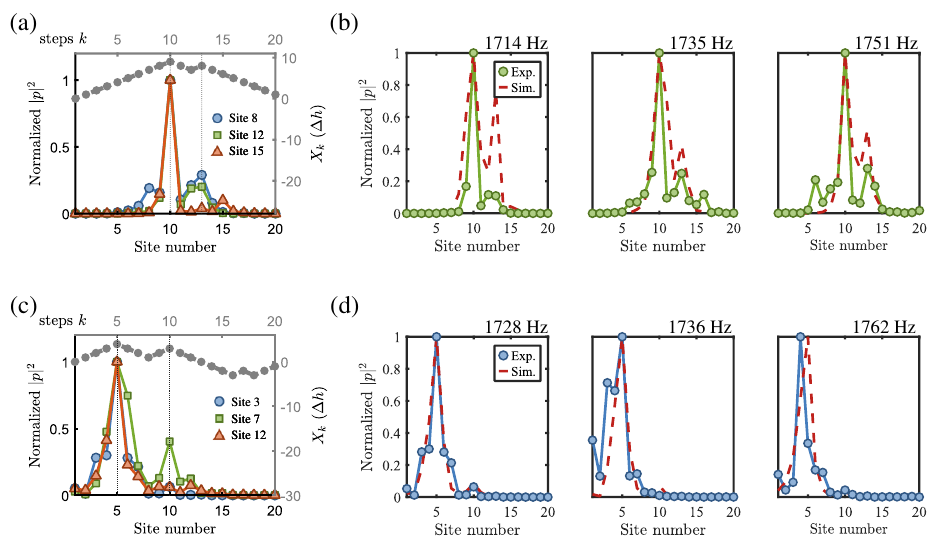}%
  \caption{\label{fig:3}Experimental results of the one-dimensional disordered Hatano--Nelson model. (a) Acoustic pressure distributions for excitation at site 8 (1721 Hz), site 12 (1715 Hz), and site 15 (1719 Hz) under a fixed disordered configuration, the upper curves correspond to the random-walk trajectories. (b) Acoustic pressure distributions at three frequencies for excitation at site 12. (c--d) Same as panels (a--b), with excitation applied at site 3 (1728 Hz), site 7 (1715 Hz), and site 12 (1736 Hz). The acoustic pressure distributions at three representative frequencies are shown.}
\end{figure*}

Intrinsic acoustic parameters extracted from experiment yield a
resonance frequency ${{f}_{0}}=1733.6~\text{Hz}$, intrinsic loss $\gamma=-4.3~\text{Hz}$,
and reciprocal coupling strength ${{\kappa}_{0}}=9.59~\text{Hz}$. Applying external gain in the
amplifier circuits produces the non-reciprocal coupling $\kappa$, with gain
levels chosen to maximize coupling while avoiding self-oscillation.
We then tune the amplification factors so that $\kappa$ remains approximately uniform across
the lattice, with a representative choice of $22-4.7i\ \text{Hz}$ (see Sec.~III of the Supplemental Material \cite{SM} for details of the experimental setup parameters, fitting method, and measured parameter values).

We now turn to the measured response of the one-dimensional disordered 
Hatano--Nelson lattice. Figures~\ref{fig:3}(a) and \ref{fig:3}(c) display the spatial sound
pressure distributions for excitations applied at selected sites under a
assigned disorder configuration, with the grey curves representing the
corresponding random-walk trajectories. The results indicate that the 
localization positions continue to follow the random-walk trajectories as 
the excitation site changes, showing that the spatial accumulation of 
acoustic energy is determined by the system's stochastic hopping sequence 
rather than the excitation site. Figures~\ref{fig:3}(b) and \ref{fig:3}(d)
show pressure distributions at three representative frequencies, with
red dashed curves denoting numerical simulations.  The results demonstrate 
good agreement between experiment and simulation: the localized peaks appear at
the extremal points of the random walk, indicating that the spatial
profiles of all eigenmodes inherit the random-walk statistics of the
hopping sequence. Although the tested sequences are random, the results
suggest that carefully designed globally reciprocal sequences could steer
energy to targeted sites. 

\begin{figure}[t]
  \centering
  \hspace*{-0.035\columnwidth}%
  \includegraphics[page=4,scale=0.85]{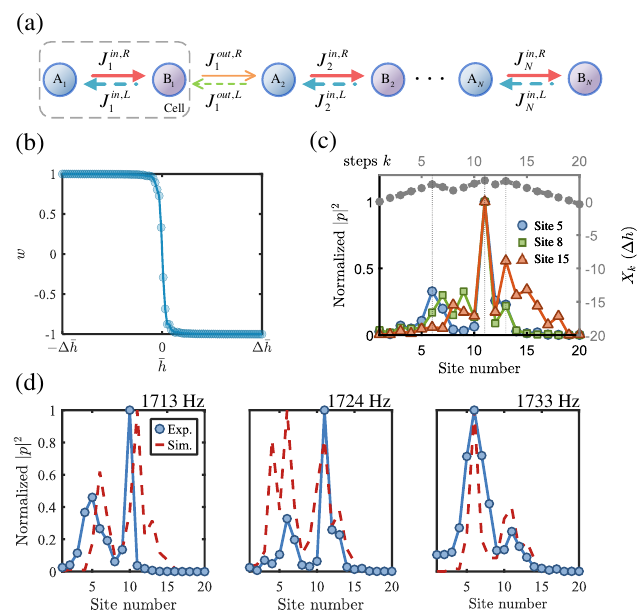}%
  \caption{\label{fig:4}Experimental results of the one-dimensional disordered SSH model. (a) Schematic diagram of the one-dimensional non-reciprocal SSH model with $N$ sites. (b) Winding numbers averaged over 50 randomly generated sequences for different mean values, with lattice size $N=1999$. (c) Acoustic pressure distributions for excitation at site 5 (1724 Hz), site 8 (1722 Hz), and site 15 (1719 Hz) under a fixed disordered configuration, the upper curves correspond to the random-walk trajectories. (d) Acoustic pressure distributions at three frequencies for excitation at site 12.}
\end{figure}

\textit{Dimerized model}---To further generalize our experiments, we constructed a one-dimensional 
lattice based on the Su--Schrieffer--Heeger (SSH)\cite{Su79,Su80} with non-reciprocal couplings, 
allowing us to explore the ENHSE behavior persists in a more 
complex lattice configuration. Each unit cell contains two sites $A_n$ and 
$B_n$ with intracell coupling $J_{n}^{\text{in},R/L}$ 
and intercell coupling  $J_{n}^{\text{out},R/L}$, each modulated by an independent
sequence defined as in Eq.~(\ref{eq:hopping}). The intracell sequence $\{h_{n}^{\text{in}}\}$
and the intercell sequence $\{h_{n}^{\text{out}}\}$ both follow a Bernoulli 
distribution, taking values $\pm \Delta h_1$ and $\pm \Delta h_2$, 
respectively. Combining the two
sequences generates the stochastic modulation that preserves the
dimerized SSH backbone while realizing a fluctuating imaginary gauge
field, as depicted in Fig.~\ref{fig:4}(a).  In this modified SSH structure, 
the intracell coupling strength remains unchanged from the previous 
experimental setup, while the intercell coupling is reduced to approximately
 $13.18-2.65i~\text{Hz}$. Averaging the real-space winding number over
50 Bernoulli realizations for a chain length of $N=1999$ [see Fig.~\ref{fig:4}(b)]
produces trends consistent with those of the Hatano--Nelson lattice,
confirming comparable point-gap topology under disorder.

The same mechanism also governs the spatial response of the system. Under a fixed disorder realization, the acoustic pressure profiles generated by excitations at different lattice sites remain pinned to the extrema of the associated random-walk trajectories, demonstrating that the spatial energy accumulation is independent of the excitation location [Fig.~\ref{fig:4}(c)]. A similar correspondence persists in the frequency domain: although different driving frequencies selectively excite distinct eigenmodes, the resulting pressure distributions consistently peak near the extremal points dictated by the underlying stochastic trajectories [Fig.~\ref{fig:4}(d)]. This indicates that the spatial structure of all eigenmodes inherits the statistical properties of the hopping disorder.
In contrast to the Hatano--Nelson lattice, the SSH geometry allows disorder to be distributed across two independent hopping channels, providing additional flexibility in realizing stochastic non-reciprocity. Nevertheless, the localization physics remains governed by the same random-walk extremal statistics. Additional experimental data are provided in Sec.~IV and Sec.~V of the Supplemental Material~\cite{SM}.

Crucially, the two independent hopping channels allow us to actively control 
the sublattice polarization of localization centers. By assigning opposite statistical 
biases to intracell and intercell couplings (for example, a negative drift for intracell 
bonds and a positive drift for intercell bonds), we maintain a vanishing global drift. As 
a result, the local extrema of $X_n$ are no longer distributed uniformly but are selectively 
biased toward one of the two sublattices. Consequently, the peaks
of the localized wave functions are predominantly localized on that sublattice.

\begin{figure}[t]
  \centering
  \includegraphics[page=5,scale=0.85]{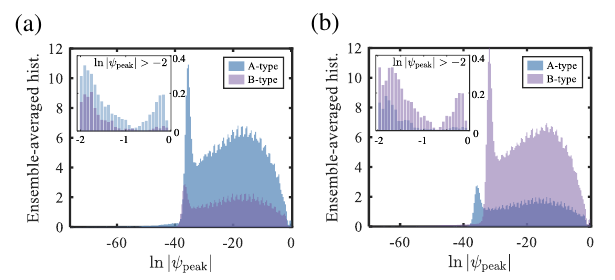}%
  \caption{Selective manipulation of localization peaks. Histograms of the logarithmic peak amplitudes $\ln|\psi_{\text{peak}}|$ for the most localized eigenstates. Blue and purple bars indicate spatial peaks pinned to sublattice A and sublattice B, respectively. The parameters are set to (a) $\Delta h_1=0.5, \Delta h_2=1.0, b_1=0.9, b_2=0.3$, and (b) $\Delta h_1=1.0, \Delta h_2=0.5, b_1=0.3, b_2=0.9$.}
  \label{fig5}
\end{figure}

To demonstrate this effect, we performed ensemble simulations across 1000 disorder realizations 
on a lattice of size $N=2000$. Setting parameters $\Delta h_1=0.5, b_1=0.9$ and $\Delta h_2=1.0, b_2=0.3$, 
where $b_1$ and $b_2$ respectively denote the probabilities of taking positive values in the corresponding 
Bernoulli distributions, induces a positive drift on intracell bonds and a compensating negative drift on 
intercell bonds. In this regime, the satellite peaks of 
the wave function (originating from the local extrema of $X_n$) are predominantly concentrated on 
sublattice A (odd sites). Consequently, the histogram of the logarithmic peak 
amplitudes [see Fig.~\ref{fig5}(a)] exhibits clear sublattice segregation dominated by contributions 
from sublattice A. To further validate this geometric mechanism, we swap the modulation parameters 
to $(\Delta h_1=1.0, b_1=0.3)$ and $(\Delta h_2=0.5, b_2=0.9)$. This substitution reverses the sign of the local 
drifts, flipping the structure of $X_n$ and shifting the principal localization maxima from sublattice A 
(odd sites) to sublattice B (even sites) [Fig.~\ref{fig5}(b)]. This deterministic inversion confirms that the 
selective localization of both primary and satellite peaks is governed by the geometry of the underlying random 
walk, providing a tunable route for steering non-Hermitian skin modes between sublattices.

\textit{Discussion}---In summary, our experimental and numerical results reveal a novel localization regime in one-dimensional non-Hermitian lattices, where bulk eigenmodes manifest as spatially discrete peaks governed by the statistics of stochastic hopping sequences. In contrast to the conventional NHSE, which drives eigenstates to accumulate at system boundaries, and distinct from Hermitian Anderson localization, which leads to a uniform suppression of transport, the ENHSE exhibits bulk peaks whose positions depend sensitively on the specific disorder realization.
In both disordered Hatano--Nelson and dimerized SSH lattices, the positions of these localized peaks are found to be locked to the extrema of the associated random-walk trajectories. This demonstrates that stochastic extreme-value fluctuations of the cumulative imaginary gauge field, rather than point-gap topology, dominate the localization mechanism in the weak-bias regime. Consistently, numerical calculations of the real-space winding number confirm that this erratic localization arises independently of topological constraints.
Furthermore, our experimental results demonstrate that site-resolved wave accumulation can be deterministically engineered by designing globally reciprocal hopping sequences, providing a viable route toward programmable wave confinement. Taken together, these results establish ENHSE as a fundamentally distinct localization phenomenon, opening new avenues for exploring the interplay among disorder, gauge fields, and lattice geometry in non-Hermitian systems.

\medskip
\noindent \textit{Acknowledgments}---This work is supported by the National Natural Science Foundation of China (Grants No. 92263208, No. 12574521, and No. 12304494), the National Key R\&D Program of China (Grant No. 2022YFA1404403), the Research Grants Council of Hong Kong SAR (Grant No. AoE/P-502/20), and the Fundamental Research Funds for the Central Universities.

\bibliography{arxiv}
\end{document}